\documentclass{amsproc}
\usepackage{graphicx}
\newtheorem{theorem}{Theorem}[section]
\newtheorem{lemma}[theorem]{Lemma}
\newtheorem{cor}[theorem]{Corollary}

\theoremstyle{definition}

\theoremstyle{remark}

\numberwithin{equation}{section}

\newcommand\C{{\mathbb C}}
\newcommand\R{{\mathbb R}}
\newcommand\cH{{\mathcal H}}
\newcommand\cM{{\mathcal M}}

\newcommand\half{{\mbox{$\frac 12$}}}
\newcommand\eps{\varepsilon}

\newcommand{\beq}{\begin{equation}}
\newcommand{\eeq}{\end{equation}}

\begin{document}

\title{Ground-State Energy of a Dilute Fermi Gas}

\thanks{Lecture given by E. Lieb at the University of Alabama,
  Birmingham 2005 International Conference on Differential Equations
  and Mathematical Physics, on the joint work of the three authors, which appeared in
Physical Review A {\bf 71},
053605  (2005) and in arXiv math-ph/0412080.}
\thanks{\copyright\, 2005 by the authors. This paper may be reproduced, in its
entirety, for non-commercial purposes.}

\author{Elliott H. Lieb}
\address{Departments of Mathematics and Physics, Jadwin Hall,
  Princeton University,\/  P.O. Box 708, Princeton, NJ 08544, USA}
\email{lieb@princeton.edu}
\thanks{The first author was supported in part by NSF Grant PHY 0139984-A03.}

\author{Robert Seiringer} \address{Department of Physics, Jadwin Hall,
  Princeton University, P.O. Box 708, Princeton, NJ 08544, USA}
\email{rseiring@princeton.edu} \thanks{The second author was supported in part by NSF Grant PHY 0353181 and by the A.P. Sloan Foundation.}

\author{Jan Philip Solovej} \address{Institute for Mathematical
  Sciences, University of Copenhagen, Universitetspar\-ken 5, DK-2100
  Copenhagen, Denmark} \email{solovej@math.ku.dk} \thanks{The third author was supported in part by EU Grant No. HPRN-CT-2002-00277, by the Danish National Research Foundation, and by grants from the Danish Research Council.}

\date{January 1, 1994 and, in revised form, June 22, 1994.}

\begin{abstract}
Recent developments in the physics of low density trapped gases make
  it worthwhile to verify old, well known results that, while
  plausible, were based on perturbation theory and assumptions about
  pseudopotentials.  We use and extend recently developed techniques
  to give a rigorous derivation of the asymptotic formula for the
  ground state energy of a dilute gas of $N$ fermions interacting with
  a short-range, positive potential of scattering length $a$. For spin
  $1/2$ fermions, this is
  $E \sim E^0 + (\hbar^2/2m) 2 \pi N \rho a$, where $E^0$ is the energy of the
  non-interacting system and $\rho$ is the density.
\end{abstract}

\maketitle

\section{Introduction}
Our goal is to find the ground state energy of a low density gas of
fermions interacting with short range pair potentials. Earlier
\cite{LY1998,LY2d,LYbham,LSSY3}, the corresponding problem for bosons had been
solved in the sense that the `well known' ancient formula $E/V=
4\pi\rho^2a$ in 3D was proved rigorously as an asymptotic formula for
small $\rho$ in the thermodynamic limit.  Here, $\rho =N/V$, where $N$ is the particle
number,  $V$ is the volume and $a$ is the scattering
length of the pair potential.

Owing to the recent interest in low density, cold gases of fermionic
atoms, the natural question arose: Is a similar formula valid for
spin-$\frac 12$ fermions? Of course there will be differences, which
we can list as follows:

\begin{enumerate}
\item We can expect that the interaction energy between pairs
  of fermions of the same spin vanishes to leading order in $\rho$.
  The Pauli principle (antisymmetry) implies that the wave function is
  tiny within the range of the pair potential.
  
\item The interaction between pairs of fermions of opposite spin
  should be the same as for bosons.  I.e., we expect this energy to be
  $4\pi a N_\uparrow N_\downarrow/V$ , where $N_\uparrow$ and
  $N_\downarrow$ are the numbers of spin up and spin down particles, 
  respectively.
  
\item A major difference from the bosonic case is that the energy we
  are trying to compute is only the second order term in the density.
  The first order term is just the kinetic energy of the ideal Fermi
  gas, which is $E_0/V= \frac 35 \left( 6\pi^2\right)^{2/3} \left\{
      (N_\uparrow /V )^{5/3} +(N_\downarrow /V)^{5/3} \right\}$.
    
  \item Finding an upper bound to the energy in the fermionic case is
    more difficult than in the bosonic case.  There, one uses the
    fact that it is {\it not} necessary to look for trial states in
    the bosonic (symmetric) sector because the bosonic energy agrees
    with the absolute minimum energy and, consequently, any
    variational state without symmetry restriction gives an upper
    bound to the bosonic ground state energy. The same is obviously
    not true for fermions, which means that we must rigorously enforce
    antisymmetry of our trial state.
  \end{enumerate}
  
  The third item gives rise to two problems that did not have to be
  faced in the bosonic case. The first is that we will have to be
  careful not to `give up' any low momentum kinetic energy in order to
  control the interaction potential. That kinetic energy is part of
  the main term in the energy and, therefore, has to be left intact.
  Since the interaction range $a$ is very small compared to the Fermi
  momentum, which is proportional to $\rho^{1/3}$, there is hope that
  in the low density regime this decomposition of the kinetic energy
  into low and high density parts can be cleanly achieved. {\it This
    is one of the main technical accomplishments of our work.}

The second main problem is closely
  related to this kinetic energy consideration. It is that the typical
  particle momentum is of the order of the Fermi momentum, which is
  not zero.  Therefore, one could question whether the `zero energy
  scattering length' contains all the necessary information about the
  interaction. That it does so is a consequence of the good separation
  of momentum scales mentioned above, but proving this fact
  mathematically requires some thought.  

In the next section we explain the problem and our results in more
detail.  The complete details can be found in \cite{LSS} and the present 
note can be regarded as a guide to the overall structure of \cite{LSS}.

\section{The Model}
Our model describes fermions interacting  through a short range
pair potential~$v$. 
The {\bf  Hamiltonian} (in units in which $\hbar = 2m =1$) is 
$$
 \boxed{\quad H_N=\sum_{i=1}^N
-\Delta_i + \sum_{1\leq i<j\leq N} v(x_i-x_j)  \quad }
$$
The nonnegative pair-potential, $v\geq0$, is radial (i.e.,
depends only on $|x_i-x_j|$) and has finite range (i.e., has compact
support).  A favorite physical example is a hard core, $v(x) =\infty$
for $|x|<a$ and $v=0 $ otherwise.

We generalize the physical problem a little by treating higher spins
than $\frac 12$, since we let the Hamiltonian $H_N$ act on the fermionic
Hilbert space with $q$ {spin states}. Spin $\frac 12$ corresponds to
$q=2$. 

The particles are restricted to a box $\Lambda=[0,L]^3$ of volume $V=L^3$.
The Hilbert space describing $N$ fermions with $q$-spin states in this box is
$$
\cH=\bigwedge^NL^2(\Lambda;\C^q),
$$
We use {Dirichlet boundary} conditions.  
Since the Hamiltonian does not depend on the spin degrees of freedom, we may 
think of the system as consisting of $q$ species of spinless fermions. 

We are interested in the {ground state energy} $
E(N,L)=\inf\,\hbox{spec}_\cH\, H_N $ in the thermodynamic limit, more
precisely the energy per unit volume defined by
$$
e(\rho)=\lim_{N,L\to\infty\atop N/L^3=\rho}\frac{E(N,L)}{L^3}.
$$
This limit is known to exist by very general arguments dating back to the 50's. 

\section{The Main Result}
\begin{theorem}[{\bf Low density asymptotics}] \label{thm:fermi} 
For $\rho\to0$ the minimum energy 
is achieved (to the two leading orders in
  $N/V$)    by having equal numbers of particles (namely $N/q$) in each spin state.  This energy is 
\begin{equation}\label{eq:main}
  \boxed{ \quad e(\rho)=\frac 35 \left( \frac{6\pi^2}{q} \right)^{2/3} \rho^{5/3} + 
  4\pi \left(1-\frac 1q\right) a_v \rho^2 + o(\rho^2), \quad }
\end{equation}
where the first term is the energy of a free Fermi gas and the second
term depends on the pair potential through its
{scattering length} $a_v$.
\end{theorem}
The {\bf scattering length} $a_v$ of the spherically symmetric
potential $v$ is defined as the constant such that the spherically
symmetric zero-energy scattering solution
$$
 -\Delta\phi+\frac12v\phi=0\ \hbox{ with }\lim_{|x|\to\infty}\phi(x)=1  
$$
satisfies
$$
 \phi(x)= 1-\frac{a_v}{|x|} 
$$
for all $x$ outside the range of $v$. 

The proof of Theorem~\ref{thm:fermi} is given in \cite{LSS}.
\medskip

{\bf Remarks:} 
(i)
The relevant dimensionless parameter is
$\rho a_v^3$.  Low density means that $\rho a_v^3\ll1$ or that $\rho\to0$, 
when $a_v$ is kept fixed.
\smallskip

(ii) For a hard core potential $a_v$ is simply
the hard core size.
\smallskip

(iii) {F}rom the definition of the scattering
length we see, by an integration by parts, that 
\begin{equation}\label{square}
\int|\nabla\phi|^2+\frac12\int v|\phi|^2=4\pi a_v.
\end{equation}
Heuristically, we may interpret this result as saying that 
for just one pair of particles in a large
volume the energy is $4\pi a_v/{\rm vol}$ per particle or altogether $8\pi a_v/{\rm vol}$. 
The number of ``interacting''
fermion pairs is $\frac{1}{2}N^2(1-\frac{1}{q})$. The interaction of
identical fermions can be ignored for low density, as explained earlier.

\section{The Bosonic Case}
The low density energy asymptotics for bosons (or even without any
symmetry restrictions, i.e., if $\cH=\bigotimes^N L^2(\R^3)$) was
proved by Lieb and Yngvason in \cite{LY1998}.  Some subsequent
developments are in \cite{LS02}--\cite{rot2}. For an up-to-date survey of rigorous
results on the Bose gas \cite{LSSY3} is recommended.

\begin{theorem}[\bf Dilute Bose gas] For $\rho\to0$
\begin{equation}\label{star}
\boxed{ \quad e^{\rm B}(\rho)=4\pi a_v \rho^2 + o(\rho^2),      \quad    }
\end{equation}
\end{theorem}
Note that the correlation term is the leading term here.\smallskip

In 1957 Dyson \cite{dyson} proved the asymptotically correct upper
bound.  This is far from trivial since one cannot use a simple product
trial function, which would give infinity in the hard core case.

Dyson also gave a lower bound for bosons, which however was not
asymptotically exact.  Dyson's lower bound relied on the following lemma, which,
in fact, was also a key ingredient in the rigorous lower bound in \cite{LY1998}.
\begin{lemma}[{\bf Dyson's Lemma}] Assume $v$ supported in $\{|x|<R_0\}$.
  Let $\theta_R$ be the characteristic function of $\{|x|<R\}$. Then for any
  positive radial function $U$ supported in $R_0\leq|x|\leq R$
  with $\int U=4\pi$ we have the operator inequality
  $$
  -\nabla\theta_R\nabla +\frac{1}{2}v\geq a_v U\ \ \ \ \mathrm{or}\ \ \ \ 
 \int \theta_R \left[ |\nabla \phi|^2 + \frac{1}{2} v |\phi|^2 \right]
\geq a_v \int U  |\phi|^2 .
 $$
\end{lemma}

Dyson's Lemma states that one may, as a lower bound, replace $v$ by a much
softer potential $w=a_v U $ with the property that $(4\pi)^{-1}\int w=(4\pi)^{-1}\int a_vU
=a_v$. However, one has to use {\it all} of the kinetic energy to do
this.

In the rigorous argument in \cite{LY1998} only {\it almost all} the kinetic
energy is used with the Dyson Lemma. 
The rest is used to carry through a perturbation argument.

As explained in the introduction we cannot afford to give up even a
fraction of the low momentum part of the kinetic energy in the
fermionic case, since it is responsible for the leading free Fermi
term.  We therefore need a generalization of Dyson's lemma, which we
describe below.

\subsection*{Early History of the Bosonic Problem}

Lenz \cite{Lenz} was the first to calculate the energy~(\ref{star}), using the
  heuristic pair argument given in Remark (iii)  after Theorem~\ref{thm:fermi}.

Bogoliubov \cite{BO} did a perturbative expansion, but got
$(8\pi)^{-1}\int v$ instead of $a_v$. He realized, of course, that this was meaningless
(e.g. for the hard core). As Bogoliubov points out in his paper,
Landau realized the connection to the scattering
  length. Namely, that $(8\pi)^{-1}\int v$ is the first term in the Born
  series for $a_v$. See also \cite{boulder}. 

Using Bogoliubov's method, the next term in the expansion beyond $4\pi
a_v \rho^2$ has been calculated to be
$$
\boxed{\quad 4\pi a_v \rho^2 \frac{128}{15\sqrt{\pi}}\sqrt{\rho a_v^3}. \quad }
$$
An important open problem is to prove this formula rigorously.  As
in the leading term, this correction depends on $v$ only through the
scattering length.

\section{The Generalized Dyson Lemma}
  
The solution to the problem that we 
cannot give up any fraction of the low momentum kinetic energy is to 
give up almost all the energy in the large momentum regime only.
Large here means momenta much greater than the Fermi momentum.  If we
give up kinetic energy corresponding to momenta greater than the Fermi
momentum only, and in such a way that as a function of momentum the
kinetic energy is still monotone, we do not affect the leading free
Fermi gas term. The following lemma, which allows us to do this, 
is the main technical advance in our work. 
  
\begin{lemma}[{\bf Generalized Dyson Lemma}]
  Assume $v$ to be supported in $\{|x|<R_0\}$.  Let $\theta_R$ be the
  characteristic function of $\{|x|<R\}$.  Let $\chi$ be a radial
  function, $0\leq \chi (p) \leq 1$, such that the Fourier transform $h(x)\equiv
  \widehat{1-\chi}(x)$ is bounded and integrable. Define
  $$
  f_R(x)= \sup_{|y|\leq R} | h(x-y) - h(x) | \quad\hbox{and}\quad
  w_R(x)= \frac{2}{\pi^2} f_R(x) \int_{\R^3} f_R(y)dy.
  $$
  Then for any $\eps>0$ and any positive radial function $U$
  supported in $R_0\leq|x|\leq R$ with $\int U=4\pi$ we have the
  operator inequality
  $$
  -\nabla \chi(p) \theta_R(x) \chi(p) \nabla + \half v(x) \geq
  (1-\eps) a_v U(x) - \frac {a_v}\eps w_R(x).
  $$
\end{lemma}

$R$ is a parameter we shall choose such that $a_v\ll R\ll\rho^{-1/3}$.
We will choose the function $\chi$ to be approximately the
characteristic function for the set of momenta greater than the Fermi
momentum, which is proportional to $\rho^{1/3}$.  This corresponds to
decomposing the kinetic energy into low and high momentum parts as shown
in the figures below.  \medskip

  \centerline{\hbox{\vbox{\hsize=0.4\hsize
        \begin{center}{$p^2(1-\chi(p))$}\\Low momentum regime
        \end{center}
        \includegraphics[height=\hsize]{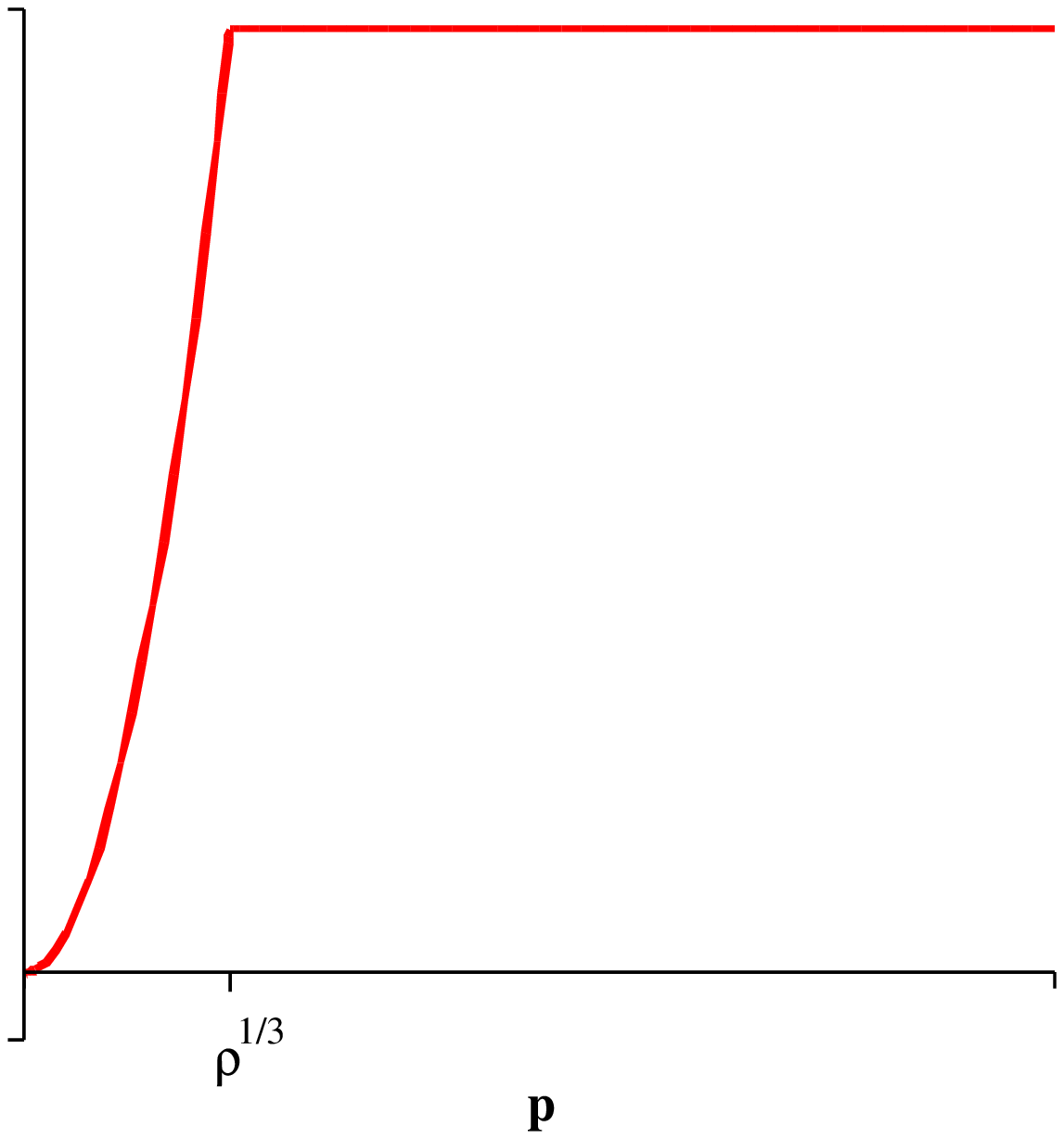}}\,\vbox{\hsize=0.4\hsize
          \begin{center}{$p^2\chi(p)$}\\High momentum regime
          \end{center}
          \includegraphics[height=\hsize]{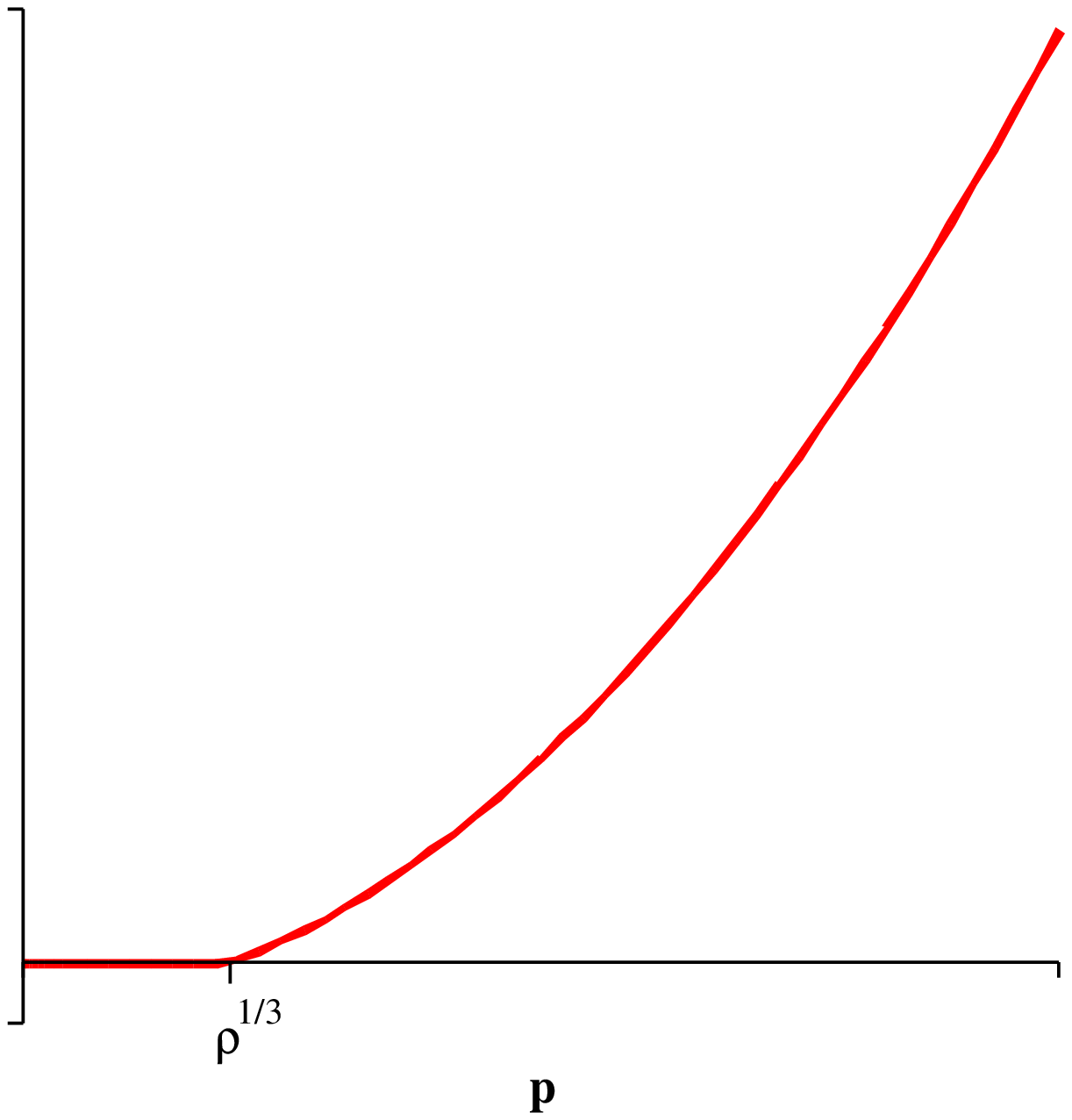}}}}

\section{A Lower Bound to the Energy}
To explain how the generalized Dyson Lemma is used to get a lower
bound to the energy we consider, for simplicity, the case $q=2$ and
write $X=(x_1,x_2\ldots, x_{N_\uparrow})$, $Y=(y_1,y_2\ldots
y_{N_\downarrow})$ for the coordinates of the two species of fermions.
Without loss of generality we may assume that
$N_\downarrow=N_\uparrow$.  The reason is that the system is $SU(2)$
symmetric and every state of total spin $S$ can be rotated to a state
of total $S_z=0$.  Nevertheless we will keep the particle numbers
arbitrary for the moment for pedagogical clarity.

We first consider the $Y$ particles as fixed.
For a lower bound we may ignore the repulsion between like fermions since $v\geq0$.
{F}rom the generalized Dyson Lemma  we obtain the following result. 
\begin{cor}[{\bf Generalized Dyson Lemma in multi-centered case}]\hfill\\
Let $R>0$ be bigger than the range of the potential $R_0$. If $|y_i-y_j|\geq 2R$
for all $i\neq j$, then
$$
-\nabla \chi(p)^2 \nabla + \half \sum_{i=1}^{N_\downarrow} v(x-y_i) \geq  \sum_{i=1}^{N_\downarrow}
\left( (1-\eps) a_v U(x-y_i) - \frac {a_v}\eps w_R(x-y_i) \right)\,.
$$
\end{cor}

To use this we must ignore $y$-particles that are closer than $2R$ apart.
This can be controlled by the following a-priori bound.
\begin{lemma}Let $I_R(Y)$ be the number of $y$'s that are
  less than $2R>0$ from their nearest neighbor. Then for any  fermionic wave function 
$\Psi(Y)$
$$
\langle\Psi,I_R\Psi\rangle\leq CT(\Psi)
R^2,\ \hbox{ where } \ T(\Psi)= \sum_{i}\langle\nabla_{y_i}\Psi,\nabla_{y_i}\Psi\rangle 
$$
and $C$ is a constant.
\end{lemma}
This lemma is an easy consequence of a theorem \cite[Theorem 5]{LYau}
relating nearest neighbor distance to kinetic energy.

For an approximate ground state $\Psi$ we can assume $T(\Psi)\leq
cN\rho^{2/3}$. Recall that we will choose $R\rho^{1/3}\ll1$, and thus
$\langle\Psi, I_R(Y)\Psi\rangle\ll N$.

The soft potential can now be studied with the remaining low momentum
kinetic energy.
$$
\sum_{i=1}^{N_\downarrow}\left[-\nabla_i(1-\chi(p))\nabla_i+{\sum}'\left( (1-\eps) a_v U(x_i-y_j) - \frac {a_v}\eps w_R(x_i-y_j) \right)\right],
$$
where $\sum'$ refers to the sum over the $N_\downarrow-I_R(Y)$ $y$-particles
that are a distance greater than $2R$ from their nearest neighbor. Since we can ignore $I_R(Y)$ the
sum of the $N_\uparrow$ lowest eigenvalues of the operator in $[\ ]$ is
approximately
$$
\frac 35 \left( 6\pi^2\right)^{2/3} {N_\uparrow}^{5/3}L^{-2}
+N_\downarrow N_\uparrow L^{-3} a_v\int U\,.
$$
We get a similar
contribution when we consider the $Y$ particles as movable and the $X$
particles as fixed.  If we also use that $\int U=4\pi$  the sum of the two contributions will 
be 
$$
\frac 35 \left( 6\pi^2\right)^{2/3} {N_\uparrow}^{5/3}L^{-2}+\frac 35 \left( 6\pi^2\right)^{2/3} {N_\downarrow}^{5/3}L^{-2}
+8\pi N_\downarrow N_\uparrow L^{-3} a_v\ .
$$
Here we see again that the minimum occurs for $N_\downarrow=N_\uparrow$, and is given by 
(\ref{eq:main}).

\section{The Upper Bound}

For an upper bound we first construct a Bijl-Dingle-Jastrow function for particles of the same spin:
\begin{equation} \label{jastrow} \nonumber
G(X)=\prod_{i<j}g(x_i-x_j)\,,
\end{equation}
where $g(x)=0$ if $|x|<R_0$, $g(x)=1$ if $|x|>s$, and $R_0\ll
s\ll \rho^{-1/3}$.
The role of $G$ is to ensure that like fermions do not interact. 

Next, we construct a Bijl-Dingle-Jastrow function for particles of opposite spin:
\begin{equation} 
F(X,Y)=\prod_{i,j}f(x_i-y_j)\,,
\end{equation} 
where $f$ (which is not normalized, since it, like $g$, 
satisfies $f(x)\to 1 $ as $|x|\to \infty$) is chosen so that
\begin{equation} \nonumber
\int|\nabla f|^2+\frac12\int
v|f|^2\approx 4\pi a_v\,.
\end{equation}
(Compare with (\ref{square}).) 
The role of $F$ is to give the correct pair energy between different
fermions.

Finally, the trial state we shall use is 
$$
\Psi(X,Y)=D(X)D(Y)G(X)G(Y)F(X,Y)\,,
$$
where $D=u_{\alpha_1}\wedge\cdots\wedge u_{\alpha_N}$ is a Slater
determinant of eigenfunctions $u_\alpha$ of the Dirichlet Laplacian.
The choice of this $\Psi$ is quite obvious. The problem is that it is
{\it not normalized}.  The normalization problem can be addressed with
the aid of the following combinatorial lemma, for which we claim no
originality.

\begin{lemma}[{\bf Key combinatorial  lemma}]
  Let $\Phi=\phi_1\wedge\cdots\wedge\phi_n$ denote a Slater
  determinant of $n$ {linearly independent} functions
  $\phi_\alpha(x)$. Let $\cM$ denote the $n\times n$ matrix
\begin{equation}\label{mmatrix}
\cM_{\alpha\beta} = \int \phi_\alpha^*(x) \phi_\beta(x)dx.
\end{equation}
\begin{itemize}
\item [(i)] The norm of $\Phi$ is given by $\langle\Phi|\Phi\rangle = \det \cM$.
\item [(ii)] For $1\leq m\leq n$, the {\underline{normalized}}
  $m$-particle density of $\Phi$ is given by
$$
(x_1,\ldots,x_m)\mapsto
\left\langle [\phi(x_1)]\wedge \cdots \wedge [\phi(x_m)] \left| \frac
    1\cM \otimes \cdots \otimes \frac 1\cM \right|
[\phi(x)]\wedge \cdots \wedge [\phi(x_m)] \right \rangle\,.
$$
\end{itemize}
Here, $[\phi(x_1)]$ denotes the $n$-dimensional vector with components
$\phi_\alpha(x)$\  .
\end{lemma}
We use this lemma, for fixed $Y$, with $\phi_\alpha(x)=u_\alpha(x)\prod_j f(x-y_j)$.
The matrix $\cM$ in (\ref{mmatrix}) then depends on $Y$ and is denoted by
$\cM(Y)$.  The following estimate shows that although the
$\phi_\alpha$ are not orthonormal, $\cM(Y)$ is close to $I$ if the
separation between the $Y$-particles is not too small.

\begin{lemma}[{\bf  Key estimate}] \label{keylemma}
Assume that $|y_i-y_j|\geq s$ for all $i\neq j$. Then
$\|I-\cM(Y)\|\to0$ as $s/a_v\to\infty$ and $N^{1/3}s/L\to 0$, uniformly in $N$ and $Y$.
\end{lemma}

Note that if $f$ were identically $1$, then $\cM$ would be equal to
the identity.  The proof of Lemma~\ref{keylemma} uses the Poincar\'e
inequality to estimate the effect of the deviation of $f$ from $1$ on
the integral in (\ref{mmatrix}).

The assumption that $|y_i-y_j|\geq s$ in Lemma~\ref{keylemma} is
satisfied since our trial function vanishes otherwise, thanks to the
property of $G$. Lemma~\ref{keylemma} is concerned only with the
effect of $F$ on the norm of $\Psi$. The effect of $G$ on the norm of
$\Psi$ has to be controlled by different means. For this purpose we
find it necessary to break space up into boxes (whose size is large
but independent of $L$) and confine the particles to these boxes,
i.e., regard particles in different boxes as independent. We refer to
\cite{LSS} for details.

\end{document}